\documentclass[%
reprint,
superscriptaddress,
 nofootinbib,
 amsmath,amssymb,
]{revtex4-2}

\usepackage{graphicx}
\usepackage{dcolumn}
\usepackage{bm}
\usepackage[version=4]{mhchem}
\usepackage{amsmath}
\usepackage{siunitx}
\usepackage{xcolor,balance} 

\usepackage[export]{adjustbox}
\begin{document}

\newcommand*\DFcomment[1]{\color{blue}{\textbf{[DF:#1]}}}
\newcommand*\SEcomment[1]{\color{yellow}{\textbf{[SE:#1]}}}

\makeatletter
    \providecommand\barcirc{\mathpalette\@barred\circ}
    \def\@barred#1#2{\ooalign{\hfil$#1-$\hfil\cr\hfil$#1#2$\hfil\cr}}
    \newcommand\stst{^{\protect\barcirc}}
    \makeatother

\preprint{APS/123-QED}

\title{`Rocket propulsion' of Janus micro-swimmers}

\author{Shaltiel Eloul}
\affiliation{Department of Chemistry, University of Cambridge, Lensfield Road, Cambridge CB2 1EW, United Kingdom} 
\author{Wilson C K Poon}
\affiliation{SUPA and School of Physics and Astronomy, The University of Edinburgh, Peter Guthrie Tait Road, Edinburgh EH9 3FD, United Kingdom}
\author{Oded Farago}
\affiliation{Department of Chemistry, University of Cambridge, Lensfield Road, Cambridge CB2 1EW, United Kingdom}
\affiliation{Biomedical Engineering Department, Ben Gurion University, Be'er Sheva 84105, Israel}
\author{Daan Frenkel}
\affiliation{Department of Chemistry, University of Cambridge, Lensfield Road, Cambridge CB2 1EW, United Kingdom}

\date{\today}
 \begin{abstract}
We report simulations of a spherical Janus particle undergoing exothermic surface reactions around one pole only. Our model excludes self-phoretic transport by design.  Nevertheless, net motion occurs from direct momentum transfer between solvent and colloid, with speed scaling as the square root of the energy released during the reaction.  We find that such propulsion is dominated by the system's short-time response, when neither the time dependence of the flow around the colloid nor the solvent compressibility can be ignored. Our simulations agree reasonably well with previous experiments.  
\end{abstract}

\maketitle

Self-propelled, or active, colloids  display novel phenomena not seen in passive suspensions such as `negative viscosity increment' and swarming collective motion \cite{poon2013physics}. The field has benefitted from the invention of synthetic micro-swimmers \cite{Howse}. Golestanian et al.~suggested that an asymmetric chemical reaction on the surface of a spherical colloid could lead to a similarly asymmetric distribution of molecular moities around the particle; the resulting concentration gradient should propel the particle by diffusiophoresis \cite{GolestanianJanus}. Self propulsion in the prototypical Janus particle, polystyrene colloids half coated with platinum \cite{Jones}, was  thought to offer a paradigmatic example, but  observation of strong salt and pH dependence pointed instead to self-electrophoresis \cite{brown_poon,ebbens2014electrokinetic}, which also propels bimetallic rod-shaped swimmers \cite{PaxtonRod}. Self-thermophoretic propulsion  is considered unlikely (but see \cite{de2013phoretic}); however, concentration gradients around an asymmetrically laser-heated particle in a near-critical binary mixture can propel micro-swimmers \cite{BechingerJanus}. The idea of non-phoretic propulsion by osmotic pressure gradients has found less acceptance~\cite{brady2011,julicher2009}, but bubble-driven propulsion of macroscopic swimmers~\cite{wang2014bubble,zhang2017self} is well established, and surface flows may propel emulsion droplets \cite{Herminghaus14}. 

Strikingly, many phoretic micro-swimmers are propelled by the decomposition of very high energy `fuel', principally hydrogen peroxide and hydrazine, two high specific impulse rocket monopropellants \cite{propellant}. Interestingly, the `detonation' of energetic molecules on a particle surface generates an impulse directly, but this has not yet been explored as a potential propulsive mechanism.

We demonstrate by mesoscopic computer simulations that such `rocket propulsion' can be practically important. We simulate an exothermic surface reaction that transfers momentum between solvent and colloid (while conserving the total system momentum), which  causes a net displacement of the colloid. A finite reaction rate then results in a relative motion of the colloid with respect to the solvent, at a speed we estimate to be non-negligible compared to self-phoretic propulsion using \ce{H2O2} as fuel. 

That there should be net propulsion is hydrodynamically unobvious. Intuitively, since the system momentum is conserved, impulsive transfer to the colloid is almost instantly cancelled by a countering fluid flow, resulting in frictional energy dissipation but no directional movement. However, this intuition neglects the finite time $\sim R ^2/\nu$ for transverse momentum to diffuse away from a colloid of radius $R $ in a fluid of kinematic viscosity $\nu$. Moreover, a third of the momentum transferred to a compressible fluid is transported away as sound \cite{Felderhof2000long} and so cannot contribute to the local flow field that slows down the colloid.  Therefore, the effect of an impulsive force on the surface is neither cancelled immediately nor locally by hydrodynamic drag forces. 

To model impulsive transport, we use dissipative particle dynamics (DPD)~\cite{hoogerbrugge1992dpd,groot1997dissipative}, which conserves momentum and thus provides a realistic  description of the hydrodynamics of a compressible fluid. 
The colloid-fluid interaction has been chosen such that there is negligible excess enthalpy or excess density of the fluid particles near the colloid. Hence, by design, self-phoretic transport should be negligible. 
We allow exothermic reactions to take place at the colloid-fluid interface, which result in a local pressure spike at the surface of the colloid.

 \begin{table}
\caption{\label{tab:table1} Simulation units, parameters and time scales }
\begin{ruledtabular}
\begin{tabular}{l|c|r}
Quantity &DPD units&Physical units\\
\hline
Mass & Fluid particle mass $m_f=1$ & $ \SI{9.76E-26}{\kilo\gram}$\\
Length& Cut off distance $r_c=1$ & \SI{6.56}{\angstrom}\\
Energy & Thermal energy $\epsilon = k_BT =1$ & \SI{4.11e-21}{\joule}\\
Speed & $(k_BT/m_f)^{0.5}=1$ & \SI{205.2}{\meter\per\second}\\
Time & $\tau = r_c (k_BT/m_f)^{-0.5} =1$ & \SI{3.2}{\pico\second}\\
Viscosity & $\eta_0 = (\epsilon/r_c^3)\tau$ = 1 & \SI{0.047}{\milli\pascal\second}\\
\hline \hline
Parameter & Value & \\
\hline
$\rho$, $\alpha,\gamma$ & 3.0, 25.0, 4.5 & \\
\hline \hline
Time scale & Value &\\
\hline
$R(=1.36) /c_s$ & $0.35$\footnote{Sound speed $c_s = \sqrt{{\rm d}p/{\rm d} \rho} \approx 4$ from the equation of state~\cite{groot1997dissipative}.} & \\
$\tau^S = m / \lambda $ &  $1.2$ & \\
$R(=1.36) ^2/\nu$ & $6.4$
\end{tabular}
\end{ruledtabular}
\end{table}

The force on DPD fluid particle $i$ is given by
\begin{equation}
\bf{f_i} =\sum_{j\neq i}{(F_{ij}^C + F_{ij}^{D} + F_{ij}^R)}\;,
\label{eq:one}
\end{equation}
where $\bf{F^{C,D,R}}$ are, respectively, conservative, dissipative and random pair forces with particle $j$. For convenience, we assume that the potential energy of interaction between two fluid particles is given by a soft  quadratic effective potential, resulting in a conservative pair force 
\begin{eqnarray}
{\bf{F_{ij}^C}} =\alpha \left(1- \frac{r_{ij}}{r_c}\right)\bf{\hat {r}_{ij}}
\end{eqnarray}
where $r_{ij}$ denotes the distance $\bf{|r_i-r_j|}$ and ${\bf \hat {r}_{ij}}={{\bf (r_i-r_j)}}/r_{ij}$ is the corresponding unit vector. The constant $\alpha$ sets the repulsion strength and mimics the compressibility of water~\cite{groot1997dissipative}. The dissipative and random forces connect via a fluctuation-dissipation relation~\cite{espanol1995statistical}:
\begin{eqnarray}
{\bf {F_{ij}^{D}}}= -\gamma \omega (r_{ij}) (\bf {v_{ij}\cdot \hat{r}_{ij})\hat{r}_{ij}}, \nonumber\\
{\bf F_{ij}^{R}}= \sqrt{2\gamma kT \omega (r_{ij})}\frac{dW_{ij}}{dt}\bf{\hat{r}_{ij}},
\label{eq:one}
\end{eqnarray}
where $\bf{v_{ij}=\left(v_i - v_j\right)}$ is the relative velocity, $\gamma$ is the friction coefficient controlling energy dissipation into the fluid, and $W_{ij}$ is a Wiener process: $\int_0^{\Delta t} dW_{ij}=\sqrt{\Delta t}\zeta_{ij}$, where $\zeta_{ij}$ is a standard Gaussian random number. 
The weight function $\omega(r)$ takes the form
\begin{eqnarray}
\omega(r_{ij}) = \left(1-\frac{r_{ij}}{r_c}\right)^2.
\end{eqnarray}

\begin{figure}[t]
  \begin{tabular}{@{}c@{}}
     \includegraphics[scale=0.17]{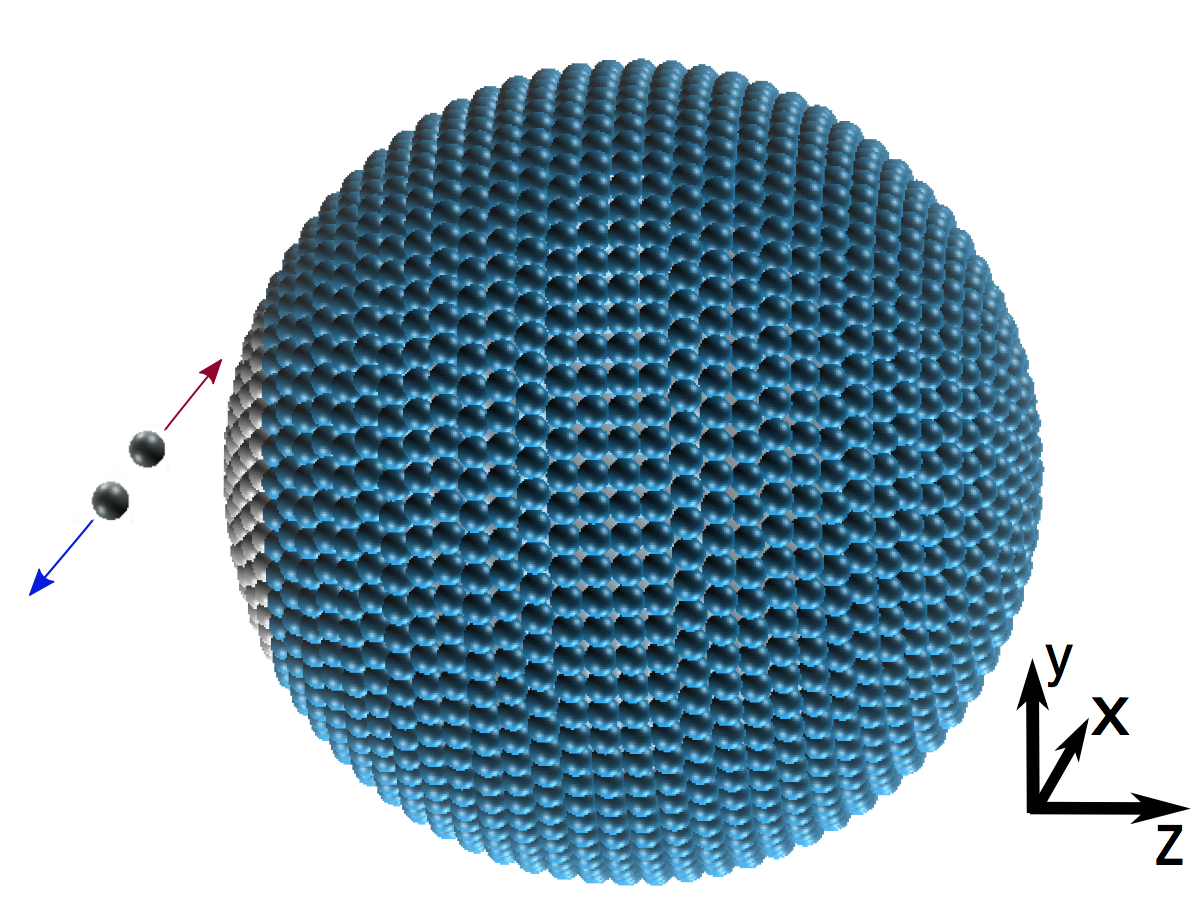}\\
    \small (a) 
     \\[\abovecaptionskip]
  \end{tabular}
  \begin{tabular}{@{}c@{}}
   \includegraphics[scale=0.15]{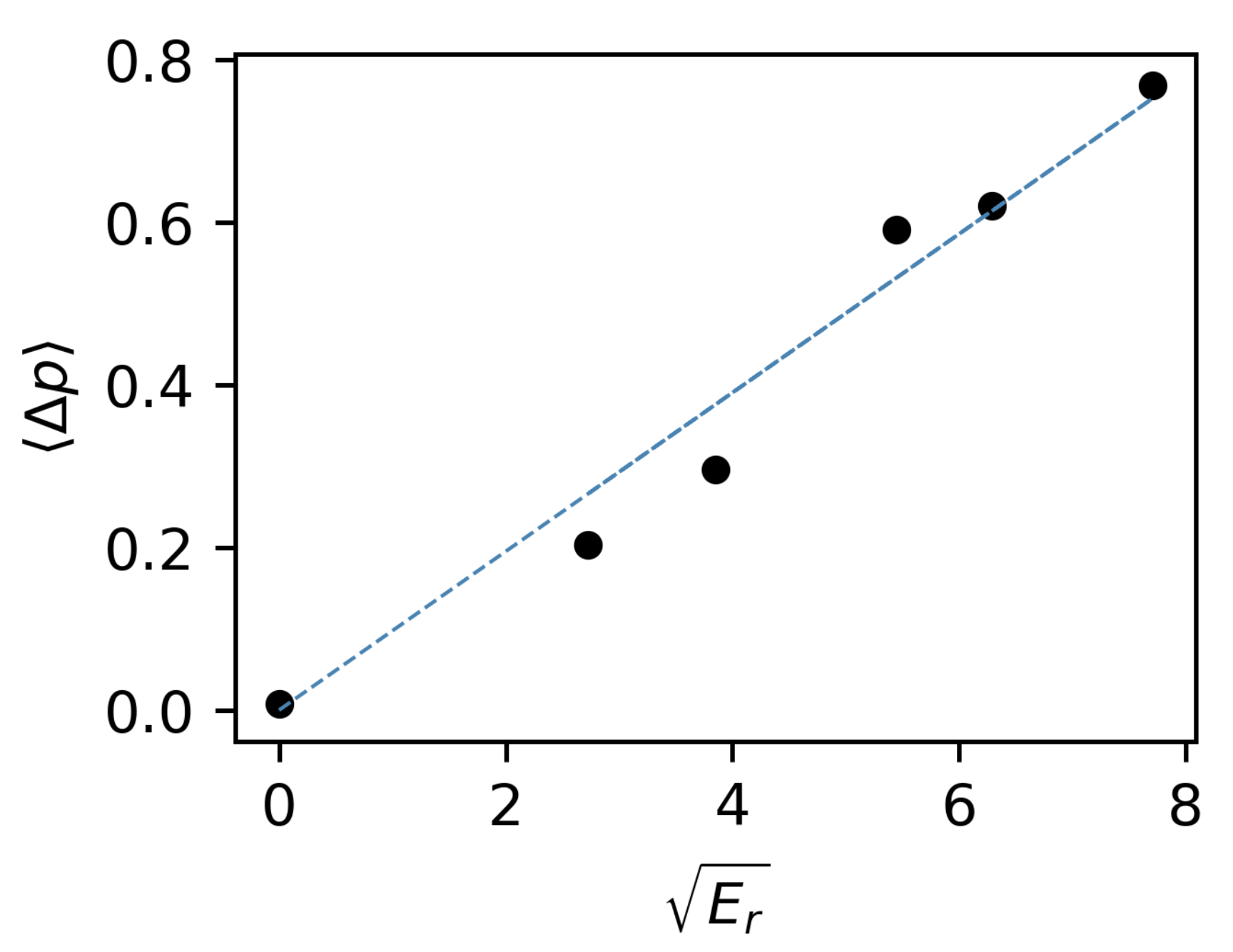}\\
   \small (b)
   \\[\abovecaptionskip]
  \end{tabular} 
  \caption{\label{fig:pvsE}
  (a) An active colloid densely covered with `frozen' DPD particles. A pair of solvent particles (black) in the vicinity of the active zone (grey) experiences a change of their velocities according to Eqns~\ref{eq:quadr0} and \ref{eq:quadr} at the time of reaction.
  (b) Average momentum transferred to the colloid from the fluid as function of square root of energy release. The line is a linear fit.}
\end{figure}

Table~\ref{tab:table1} summarises our units and parameters. The cut-off distance $r_c$ is our length unit. Each DPD particle has mass $m_f$ chosen to be the mass of three molecules with molecular weight 19.8, corresponding to a 10~wt.\% \ce{H_2O_2} solution for comparability with experiments. To reproduce the mass density of this solution, \SI{1040}{\kilo\gram\per\cubic\meter}~\cite{h2o2density}, we use a DPD density of $\rho  r_c^3=3.0$. To reproduce the compressibility of water at room temperature, $\alpha=25$ and $\gamma=4.5$~\cite{groot1997dissipative}. The thermal energy $\epsilon = k_B T$ is our energy unit, and $m_f$ is our mass unit, so that our time unit is $r_c\sqrt{m_f /\epsilon}$. Using a previous result~\cite{groot1997dissipative}, we estimate a shear viscosity of $\eta = 0.96$ in DPD units, corresponding to \SI{4.7e-5}{\pascal\second} in real units. This is $\sim 20$ times lower than that of water; we later correct for when comparing with experiments. The fluid equations of motion were integrated using the modified velocity-Verlet algorithm~\cite{groot1997dissipative} with a reduced time step of $5 \times 10^{-3}$. 

We model the Janus colloid using a  dense spherical layer of `frozen' DPD particles, Fig.~\ref{fig:pvsE}a. Its radius $R$ (= 2.3 $\equiv \SI{1.5}{\nano\meter}$ unless otherwise stated) is defined as the distance between the center of the colloid  and the centers of the surface particles. The surface-particle density is chosen to be high enough to suppress penetration of fluid particles during the simulation. Moreover, we found that for surface densities $\rho_s\gtrsim25$,  the speed of the reaction-driven colloid was insensitive to $\rho_s$. With enough fluid particles surrounding the colloid, it is sufficient not to include dissipative and random forces between the `frozen' surface particles and fluid particles, and take this interaction as repulsive only.  This has only a small effect on the hydrodynamic boundary conditions  (the  surface of the colloid is still fairly rough) and does not change our conclusions qualitatively. 
Since the fluid particles interact with the colloid through conservative forces, the colloid acquires the temperature of the DPD fluid in the absence of chemical reactions. We freeze out the rotational motion of the colloid, which only acts on longer time scales. 
The total force on the center-of-mass of the colloid is the sum of all (repulsive) forces between the `frozen' particles on its surface and the neighbouring fluid particles. The equations of motion of the colloid are also solved using the velocity-Verlet algorithm. 

To model a Janus swimmer, we chose parameters appropriate for the reaction $\ce{2H_2O_2 -> O_2 + 2H_2O}$ on a \ce{Pt} surface with standard enthalpy $\Delta H\stst = \SI{1.017}{\electronvolt} \equiv 39.6k_BT$ at room temperature. A reaction is modelled by increasing the kinetic energy of a pair of neighboring DPD particles close to the catalytic surface by $\Delta H\stst$, conserving momentum and leaving all species unchanged. For simplicity but without loss of generality, we constrain the active zone for reactions to a small area on the particle surface (grey patch in Fig.~\ref{fig:pvsE}a). Reactions occur at a rate, or frequency, $f$ (in inverse time units). 

The DPD thermostat acts as a local energy sink and reduces the efficiency of momentum transfer due to reaction.  For comparison, we carried out a simulation  without frictional forces. We found that  the speed of the colloid increased by a factor of $\sim 2.2$. We used a thermostat simply to suppress heating of the fluid and to minimise possible temperature gradients along the colloidal surface. The latter point is probably less important as we designed our model such that the excess surface enthalpy (and hence any thermophoresis) is minimized.
Diffusiophoresis can also be ignored because (again by design) the reactants and products are identical and so have the same interaction with the colloidal surface.

As a result of a reaction event, the kinetic energy of a pair of neighboring particles is increased by an amount $E_r=\Delta H^{\stst}$. Energy conservation  then leads to:
\begin{equation}
{
{({\bf v_{a}} +\Delta {\bf v_a)}}^2+({{\bf v_{b}} +\Delta {\bf v_b)}}^2  = {\bf{v}_a}^2+{\bf{v}_b}^2 }+ \frac{2}{m_f}E_r \;,
\end{equation}
where $\bf{v_a}$ and $\bf{v_b}$ are the particle velocities before the reaction, which are changed by $\Delta \bf{v}_a$ and $\Delta \bf{v}_b$ due to energy injection. Momentum conservation requires: 
\begin{equation}\label{eq:quadr0}
\Delta {\bf v_a} = - \Delta {\bf v_b}\;,  
\end{equation}
so that 
\begin{equation}\label{eq:quadr}
{
\Delta {\bf v_a}^2+\Delta {\bf v_a} \cdot ({\bf v_{a}}-{\bf v_{b}}) =\frac{E_r}{m_f}.
}
\end{equation}
The direction of $\Delta {\bf v_a}$ is randomly chosen from a uniform distribution; its magnitude follows from Eqn~\ref{eq:quadr}.

We measure an average net momentum transfer to the colloid as a result of near-surface reactions, $\langle \Delta p\rangle$, which scales as $E_r^{0.5}$, Fig.~\ref{fig:pvsE}b. This is expected on dimensional grounds if this momentum transfer is indeed due to `rocket propulsion'. The observed scaling rules out self-thermophoresis, for which $\langle \Delta p \rangle \sim E_r$, and self-diffusiophoresis, for which $\langle \Delta p \rangle \sim E_r^0$. Note that only a fraction of $\Delta H\stst$ is converted into momentum of the colloid.  The precise fraction will depend on details, including the model parameters that determine friction.

 \begin{figure}[t]
\includegraphics[scale=0.27]{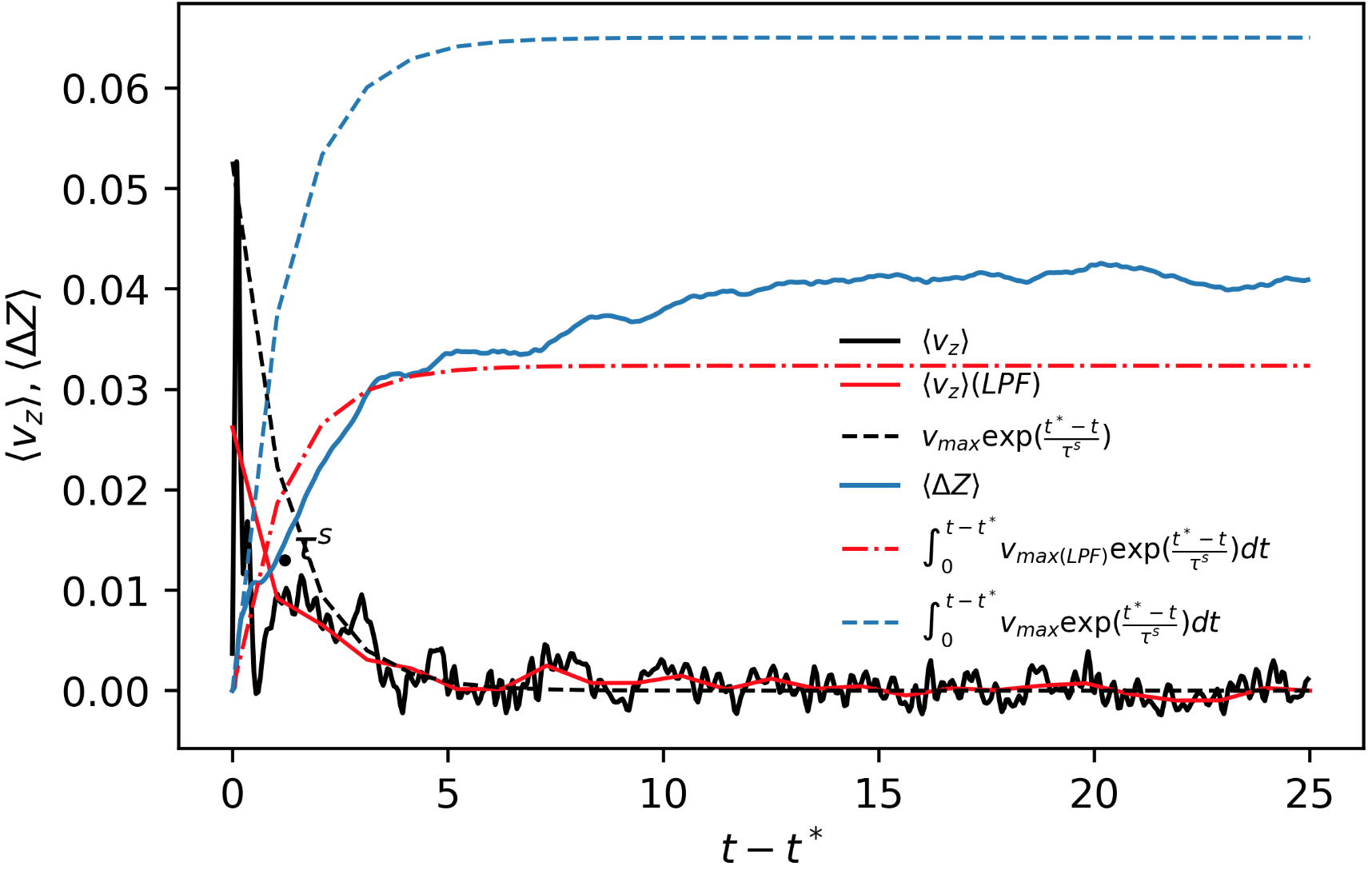}
\caption{\label{fig:react_profile} Average transient profile of the velocity decay (black curve) and total displacement (blue curve) of an active colloid ($R =1.36$) as function of time ($t$) from the time of reaction ($t^*$) for $E_r = 39.6k_BT$.  The momentum loss and the total displacement after the reaction impulse are also compared to Stokes friction velocity damping (black dashed curve) and the time integral (blue dashed curve). Additional momentum decay and displacement curves (red) shows the result of low pass filtering (LPF) to remove fast oscillation.}
\end{figure}

The finite momentum transfer imparts a net transient velocity to the colloid along its polar, or $z$, axis, which points directly away from the active patch. Fig.~\ref{fig:react_profile} (black line) shows this velocity averaged over $10^4$ independent reaction events, $\langle v_z \rangle$, at a low reaction rate ($f = 0.04$) as a function time elapsed since the reaction, $t - t^*$. Despite the noise, non-monotonicity is apparent at short times, probably due to the compressibility of the fluid. We average out such short-time non-monotonicity as well as the noise using a low-pass filter to give $\langle v_z \rangle$(LPF) (red line). Neither the raw nor the low-pass filtered data follow the single exponential (black dashed) from an initial $\langle v_0 \rangle$ predicted by Stokes Law:
\begin{equation}\label{eq:veldamp}
{
\langle v_z \rangle=\langle v_0 \rangle\exp{\left[-\left(\frac{t-t^*}{\tau^S} \right)\right]},
}
\end{equation}
where $\tau^S=m /\lambda  \approx 1.2$ (with $\lambda $ the Stokes drag coefficient). The actual decay of $\langle v_z \rangle$ is substantially faster, because our `detonation' {\it ansatz} generates a force {\it dipole} rather than a monopole in the fluid.

\begin{figure}[t]
  \centering
  \begin{tabular}{c}
     \includegraphics[scale=0.21]{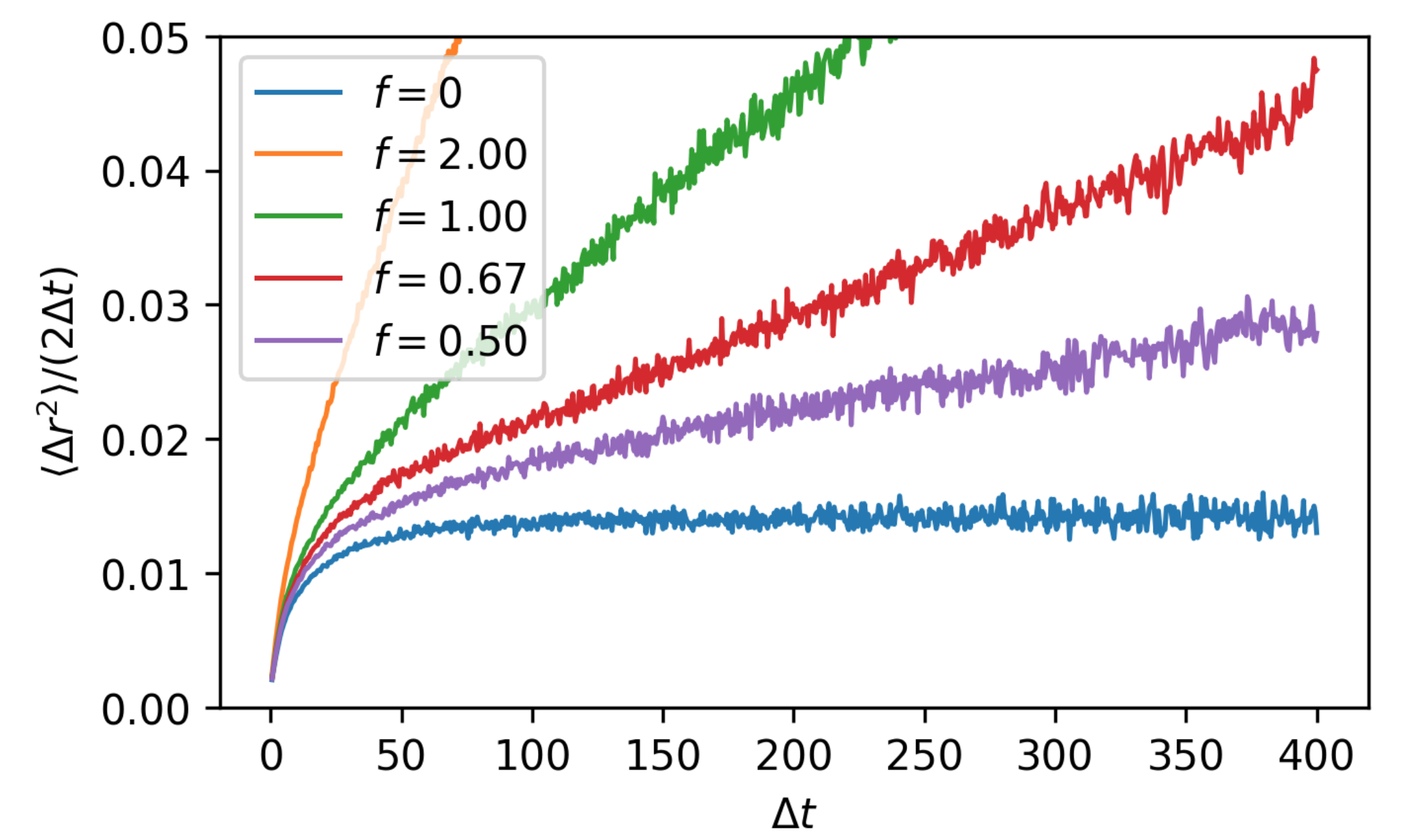}\\
   \small (a)\\
     [\abovecaptionskip]
  \end{tabular}
  \begin{tabular}{@{}c@{}}
  \includegraphics[scale=0.22]{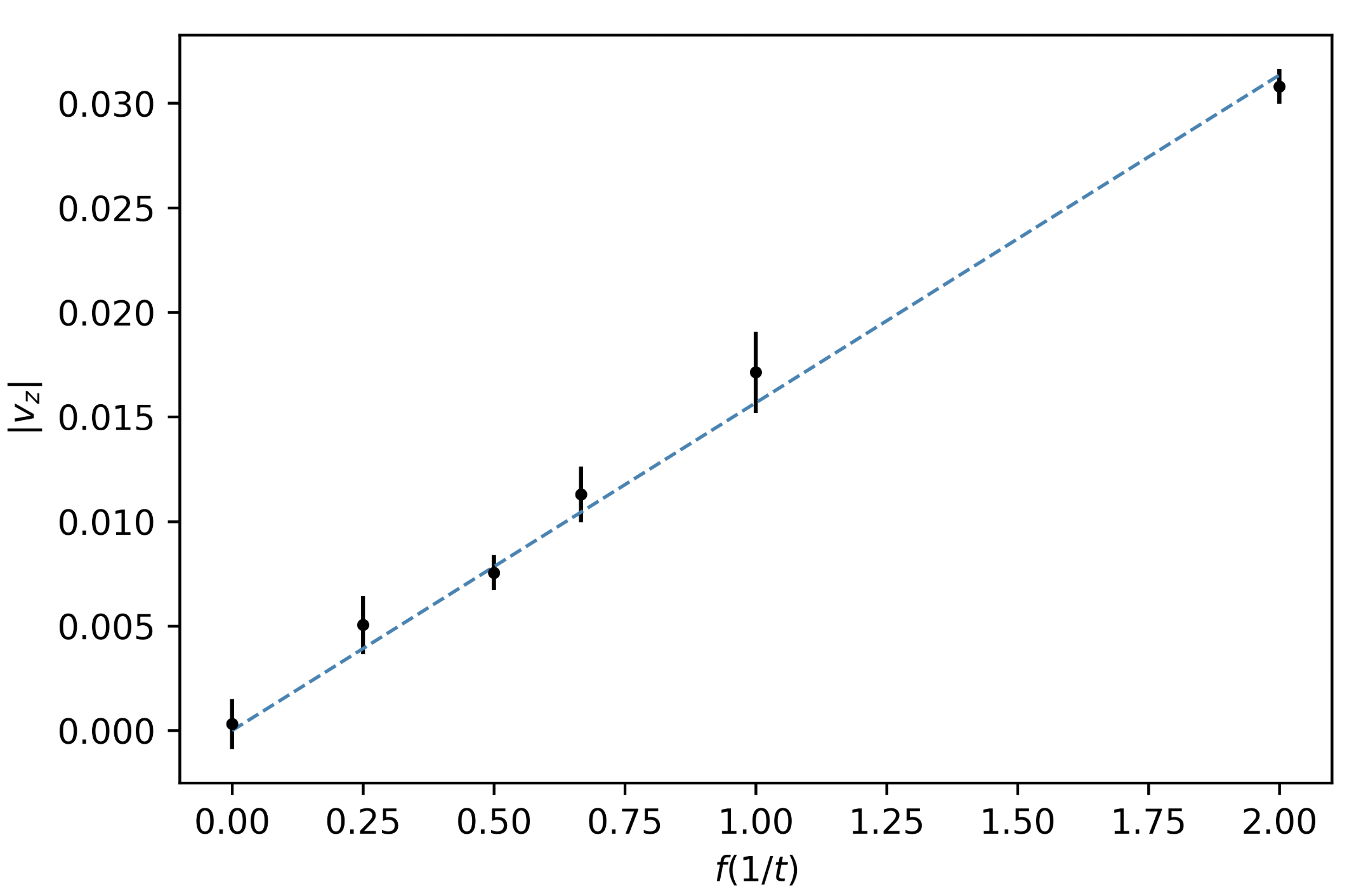}\\
      \small (b)
      \\[\abovecaptionskip]
  \end{tabular} 
  \caption{\label{fig:active_msdondt} (a) $\langle \Delta r^2 \rangle/2\Delta t$  as function of $\Delta t$ for various reaction freq. (b) Drift velocity in $z$ direction as function of reaction frequency (linear fit shown as dashed line). The average velocity and the standard deviations are extracted from total of 15,000 reactions. In both parts, $E_r = 39.6k_BT$.}
\end{figure}

Integrating the average transient velocity gives the average displacement of the colloid in response to a reaction event, $\langle \Delta z \rangle$, Fig.~\ref{fig:react_profile} (blue line). There is a rapid rise in displacement at short times. One might expect that this rapid rise should saturate beyond $t - t^* \lesssim R ^2/\nu \approx 6.4$. This is indeed what we see for the displacement from integrating the smoothed data (red dot-dashed). The rise in the actual displacement (blue) does slow down around this time, but continues to rise to saturate at a value that is about a third higher. This may be related to the `long-time tail' in the velocity autocorrelation function, although our statistics are not good enough to quantify this effect. (Note that the long-time tail in our system will not follow the well-known $t^{-3/2}$ form because, once again, the chemical reaction results in a force dipole.)

Fig.~\ref{fig:active_msdondt} shows the accumulated mean-squared displacement, $\langle \Delta r^2 \rangle$, of the colloid due to a succession of chemical reactions as a function of the elapsed time, $\Delta t$, at different reaction frequencies, $f$. For $f = 0$, $\langle \Delta r^2 \rangle/2\Delta t$ saturates in the long-time limit: the passive colloid is diffusive, and its diffusivity satisfies the Stokes-Einstein relation (see Supplementary Information). At all $f>0$, however, $\langle \Delta r^2 \rangle/2\Delta t$ no longer saturates with time, but asymptotes to a  linear regime, indicative of ballistic motion at a constant drift speed. 

\begin{figure}[t]
\includegraphics[scale=0.25,left]{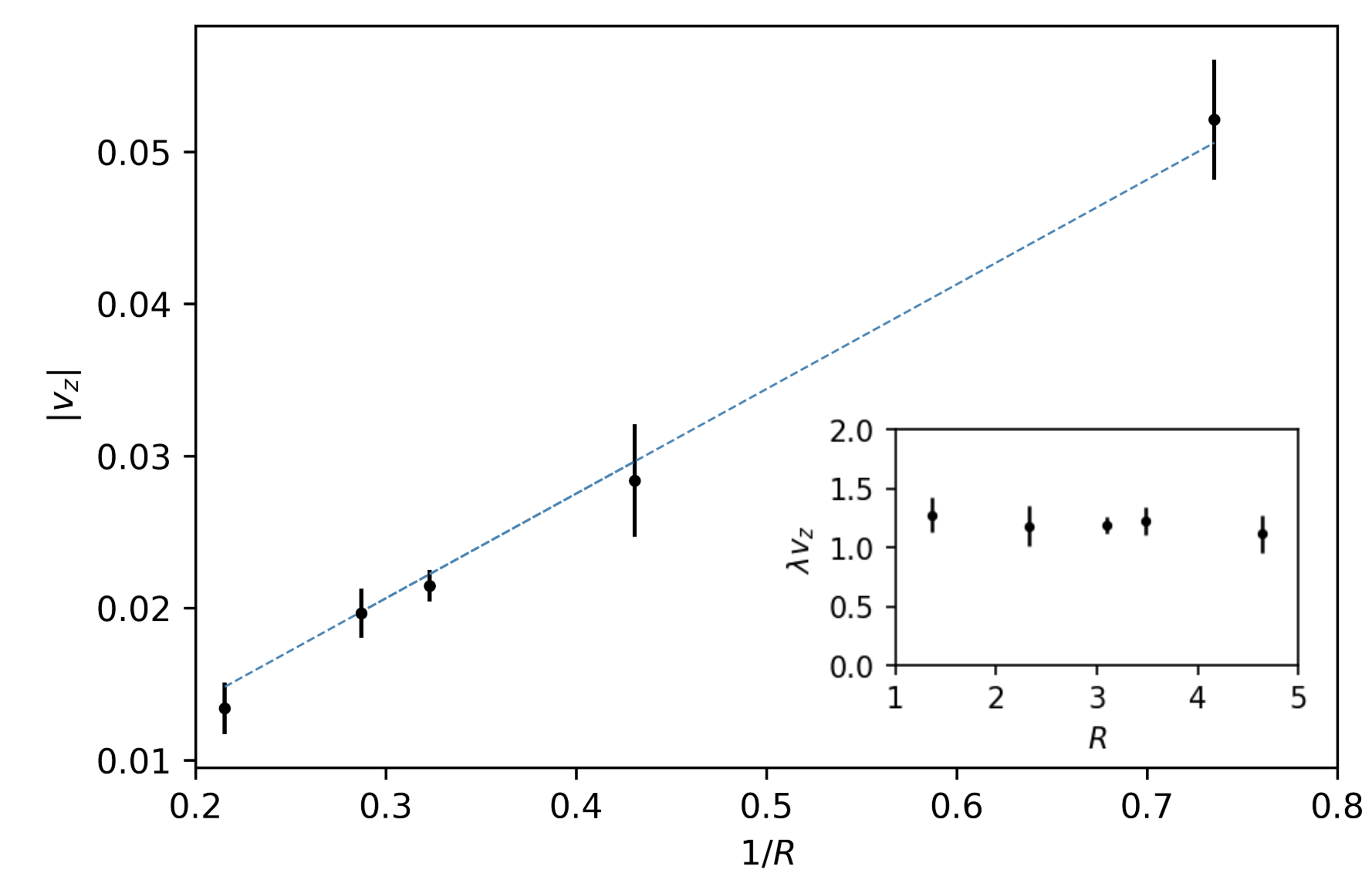}
\caption{\label{fig:RVzvsR} Averaged velocity (Sample average and standard deviations of trajectories contain a total of $10^4$ reactions in each data point) as function of the colloidal radius for the case of $f=2$ (The fitted linear curve shown in blue dashed line).   Inset: $\lambda v_z$ as a function of $R$.}
\end{figure}

We verified that the drift speed along the $x$ and $y$ axes imparted by reactions averaged to zero. This average drift speed along the $z$, or polar, axis of the colloid, $|v_z|$, increases linearly with $f$ as shown in Fig.~\ref{fig:active_msdondt}b. Thus, the effects of successive reaction events are simply additive, and $|v_z|$ can be related to the average momentum transfer per reaction $\langle \Delta p \rangle$, the  reaction frequency $f$ and the hydrodynamic friction coefficient $\lambda$ of the colloid by
\begin{equation}\label{eq:v_drift}
{
|v_z| = \frac{\langle \Delta p \rangle}{\lambda }f = \left(\frac{\langle \Delta p \rangle}{6\pi \eta }\right)\frac{f}{R }.
}
\end{equation}
The second equality, which follows from $\lambda = 6\pi\eta R$, predicts that $|v_z| \propto R^{-1}$. Fig.~\ref{fig:RVzvsR} shows that, to within the statistical error, this is indeed the case for simulations over the range $1.36 \leq R \leq 4.6$ at constant $f$ (where the ratio of the colloidal radius to the box diameter is kept approximately constant).  The fact that $|v_z| \propto R^{-1}$ implies that $\langle \Delta p \rangle$ (the average momentum transferred to the colloid) is independent of the the colloidal radius. Similarly, the inset of Fig.~\ref{fig:RVzvsR} shows that $\lambda v_z$, {\em i.e.} the average momentum transfer per unit time, is effectively  independent of $R$ over this range of radii. 

We now compare the prediction of Eqn~\ref{eq:v_drift} to previous experimental data~\cite{brown_poon} for a \SI{2}{\micro\meter}-diameter Janus particle half coated with Platinum in 10~wt.\% aqueous \ce{H2O2} with a measured reaction rate of \SI{8e10}{\per\second}, or $R = 1524$ and $f=0.24$ in DPD units (cf.~Table~\ref{tab:table1}). Both the linear fits in Fig.~\ref{fig:active_msdondt}b and Fig.~\ref{fig:RVzvsR} along with the DPD viscosity of $\eta = 0.96$ give, via Eqn~\ref{eq:v_drift} the momentum transfer per reaction, $\langle \Delta p \rangle=0.62$, which, as we have seen, is nearly $R$-independent in our simulation range. We extrapolate this to larger radii. Additionally, the reaction is confined to a small area around one pole in our simulations, Fig.~\ref{fig:pvsE}a. A hemispherical coating will reduce the effective momentum transfer by a factor of two, giving $\langle \Delta p \rangle = 0.31$ for Janus particles. We therefore predict a drift speed of $2.64\times10^{-6}$ in DPD units. This is an overestimate, because the viscosity of our DPD fluid is 20 times lower than that of water, so that we finally predict a drift speed of $1.32 \times 10^{-7}$ in DPD units, or $\approx \SI{27}{\micro\meter\per\second}$. 

\begin{figure}[t]
\includegraphics[scale=0.25,left]{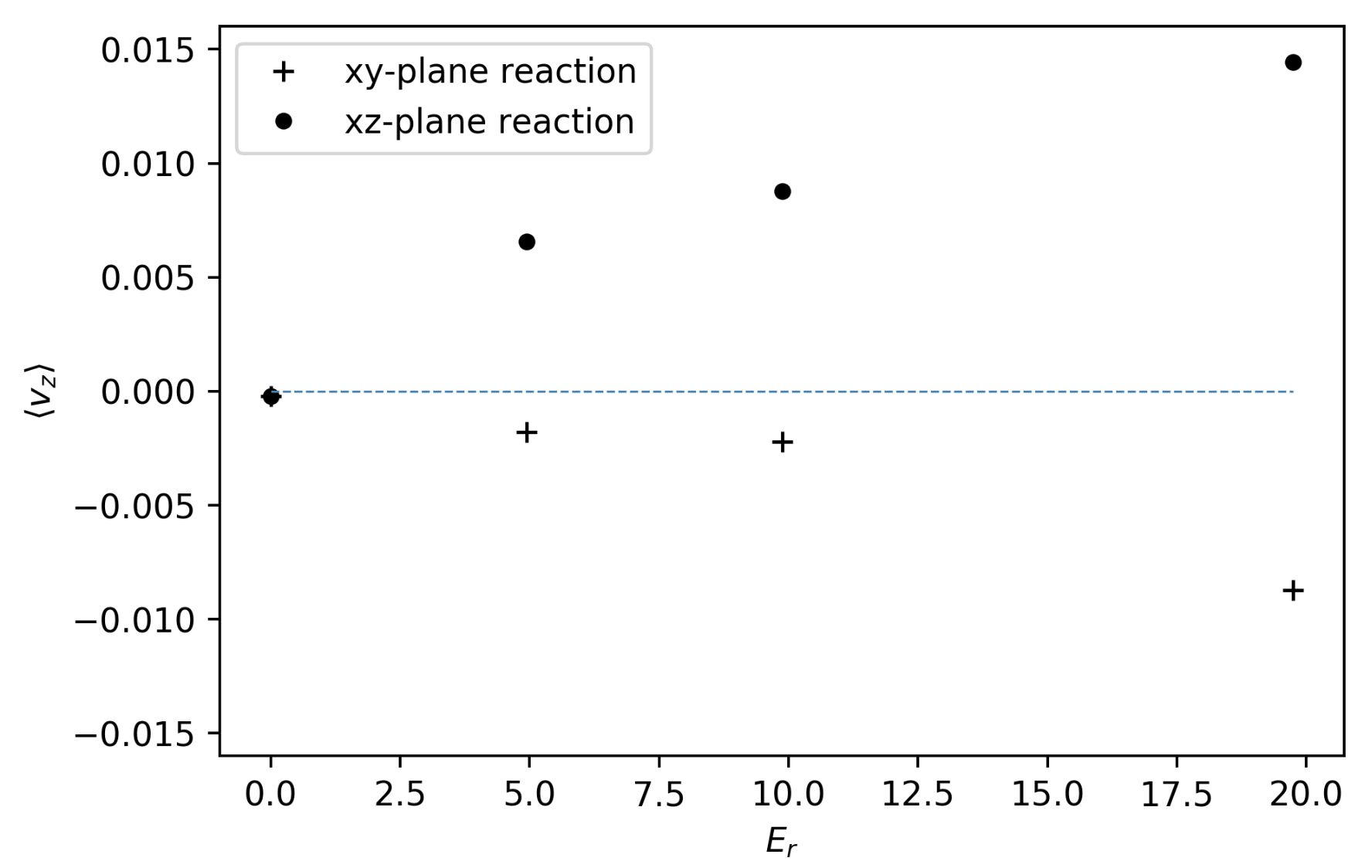}
\caption{\label{fig:vvsEplane} Colloidal drift velocity as function of reaction energy release $E_r$ for $R=3.2$ and. $f = 2$.  The reaction products are constrained to lie in the $xy$ (+) and $xz$ ($\bullet$) planes.}
\end{figure}

This is more than twice the observed value of $11 \pm \SI{6}{\micro\meter\per\second}$~\cite{brown_poon}. Nevertheless, the good order of magnitude agreement suggests strongly that impulsive propulsion cannot be ignored in real systems, especially because it is a direct manifestation of momentum conservation, and so cannot be `designed out' in the way that we have removed diffusiophoresis and thermophoresis in our simulations. That our prediction is off by a numerical factor simply reflects the crudeness of our model. However, that we have an {\it over}estimate merits further investigation. 

Our `detonation {\it ansatz}' assumes that the direction of the relative motion of reaction products is randomly distributed relative to the particle surface. However, if we constrain the initial motion of the reaction products to the $xz$ or $xy$ plane (cf.~Fig.~\ref{fig:pvsE}), the resulting $z$-drift velocity is reduced, and may even change sign, Fig.~\ref{fig:vvsEplane}. Adsorption of reactant molecules on the catalytic surface in different orientations may impose such constraints, reducing the drift speed. Moreover, real catalytic coatings are rough, so that reactions may occur in solvent trapped in pockets. In a thought experiment in which reaction occur entirely within a spherical pocket inside the colloid, the net average momentum transfer will in fact be zero, suggesting a possible test of our model. 

Further experimental confrontation is suggested by the near-$R$-independence of $\langle \Delta p \rangle$, Fig.~\ref{fig:RVzvsR}. This allows us to predict how the speed of a `rocket propelled' Janus particle should depend on radius using Eqn~\ref{eq:v_drift}. In a system where the reaction on the colloid surface is diffusion controlled, we expect $f \propto R$, and Eqn~\ref{eq:v_drift} predicts that $|v_z|$ should be independent of radius. On the other hand, if the reaction is rate limited, $f \propto R^2$, so that $|v_z| \propto R$.

In sum, we find that energy release during an exothermic reaction can propel a colloid due to impulsive momentum transfer, which can never be `turned off' in experimental systems of this kind. Our model is undoubtedly over-simplified and there are many other factors that may affect the efficiency of momentum transfer between solvent and colloid, such as the details of catalytic decomposition and surface topography. Nevertheless, we find the magnitude of speeds attainable means that this mechanism can seldom, if ever, be ignored as one of the propulsion mechanisms of real-life micron-sized Janus particles in which phoretic mechanisms also operate.

\begin{acknowledgments}
 The Cambridge work was partially funded by the  Horizon 2020 program through 766972-FET-OPEN-NANOPHLOW. WCKP was funded by ERC Advanced Grant ERC-2013-AdG 340877-PHYSAP. We thank Alexander Morozov for enlightening discussions.
\end{acknowledgments}

\balance

\bibliography{active_colloid}

\providecommand{\noopsort}[1]{}\providecommand{\singleletter}[1]{#1}%
\begin{thebibliography}{21}%
\makeatletter
\providecommand \@ifxundefined [1]{%
 \@ifx{#1\undefined}
}%
\providecommand \@ifnum [1]{%
 \ifnum #1\expandafter \@firstoftwo
 \else \expandafter \@secondoftwo
 \fi
}%
\providecommand \@ifx [1]{%
 \ifx #1\expandafter \@firstoftwo
 \else \expandafter \@secondoftwo
 \fi
}%
\providecommand \natexlab [1]{#1}%
\providecommand \enquote  [1]{``#1''}%
\providecommand \bibnamefont  [1]{#1}%
\providecommand \bibfnamefont [1]{#1}%
\providecommand \citenamefont [1]{#1}%
\providecommand \href@noop [0]{\@secondoftwo}%
\providecommand \href [0]{\begingroup \@sanitize@url \@href}%
\providecommand \@href[1]{\@@startlink{#1}\@@href}%
\providecommand \@@href[1]{\endgroup#1\@@endlink}%
\providecommand \@sanitize@url [0]{\catcode `\\12\catcode `\$12\catcode
  `\&12\catcode `\#12\catcode `\^12\catcode `\_12\catcode `\%12\relax}%
\providecommand \@@startlink[1]{}%
\providecommand \@@endlink[0]{}%
\providecommand \url  [0]{\begingroup\@sanitize@url \@url }%
\providecommand \@url [1]{\endgroup\@href {#1}{\urlprefix }}%
\providecommand \urlprefix  [0]{URL }%
\providecommand \Eprint [0]{\href }%
\providecommand \doibase [0]{https://doi.org/}%
\providecommand \selectlanguage [0]{\@gobble}%
\providecommand \bibinfo  [0]{\@secondoftwo}%
\providecommand \bibfield  [0]{\@secondoftwo}%
\providecommand \translation [1]{[#1]}%
\providecommand \BibitemOpen [0]{}%
\providecommand \bibitemStop [0]{}%
\providecommand \bibitemNoStop [0]{.\EOS\space}%
\providecommand \EOS [0]{\spacefactor3000\relax}%
\providecommand \BibitemShut  [1]{\csname bibitem#1\endcsname}%
\let\auto@bib@innerbib\@empty
\bibitem [{\citenamefont {Poon}(2013)}]{poon2013physics}%
  \BibitemOpen
  \bibfield  {author} {\bibinfo {author} {\bibfnamefont {W.~C.~K.}\
  \bibnamefont {Poon}},\ }\bibfield  {title} {\bibinfo {title} {From
  \emph{Clarkia} to \emph{Escherichia} and {J}anus: The physics of natural and
  synthetic active colloids},\ }in\ \href@noop {} {\emph {\bibinfo {booktitle}
  {Physics of Complex Colloids}}},\ \bibinfo {editor} {edited by\ \bibinfo
  {editor} {\bibfnamefont {C.}~\bibnamefont {Bechinger}}, \bibinfo {editor}
  {\bibfnamefont {F.}~\bibnamefont {Sciortino}},\ and\ \bibinfo {editor}
  {\bibfnamefont {P.}~\bibnamefont {Ziherl}}}\ (\bibinfo  {publisher}
  {Societ\`a Italiana di Fisica},\ \bibinfo {address} {Bologna},\ \bibinfo
  {year} {2013})\ pp.\ \bibinfo {pages} {317--386}\BibitemShut {NoStop}%
\bibitem [{\citenamefont {Ebbens}\ and\ \citenamefont {Howse}(2010)}]{Howse}%
  \BibitemOpen
  \bibfield  {author} {\bibinfo {author} {\bibfnamefont {S.~J.}\ \bibnamefont
  {Ebbens}}\ and\ \bibinfo {author} {\bibfnamefont {J.~R.}\ \bibnamefont
  {Howse}},\ }\bibfield  {title} {\bibinfo {title} {In pursuit of propulsion at
  the nanoscale},\ }\href@noop {} {\bibfield  {journal} {\bibinfo  {journal}
  {Soft Matter}\ }\textbf {\bibinfo {volume} {6}},\ \bibinfo {pages} {726}
  (\bibinfo {year} {2010})}\BibitemShut {NoStop}%
\bibitem [{\citenamefont {Golestanian}\ \emph {et~al.}(2005)\citenamefont
  {Golestanian}, \citenamefont {Liverpool},\ and\ \citenamefont
  {Ajdari}}]{GolestanianJanus}%
  \BibitemOpen
  \bibfield  {author} {\bibinfo {author} {\bibfnamefont {R.}~\bibnamefont
  {Golestanian}}, \bibinfo {author} {\bibfnamefont {T.}~\bibnamefont
  {Liverpool}},\ and\ \bibinfo {author} {\bibfnamefont {A.}~\bibnamefont
  {Ajdari}},\ }\bibfield  {title} {\bibinfo {title} {Propulsion of a molecular
  machine by asymmetric distribution of reaction products},\ }\href@noop {}
  {\bibfield  {journal} {\bibinfo  {journal} {Phys. Rev. Lett.}\ }\textbf
  {\bibinfo {volume} {94}},\ \bibinfo {pages} {220801} (\bibinfo {year}
  {2005})}\BibitemShut {NoStop}%
\bibitem [{\citenamefont {Howse}\ \emph {et~al.}(2007)\citenamefont {Howse},
  \citenamefont {Jones}, \citenamefont {Ryan}, \citenamefont {Gough},
  \citenamefont {Vafabakhsh},\ and\ \citenamefont {Golestanian}}]{Jones}%
  \BibitemOpen
  \bibfield  {author} {\bibinfo {author} {\bibfnamefont {J.}~\bibnamefont
  {Howse}}, \bibinfo {author} {\bibfnamefont {R.}~\bibnamefont {Jones}},
  \bibinfo {author} {\bibfnamefont {A.}~\bibnamefont {Ryan}}, \bibinfo {author}
  {\bibfnamefont {T.}~\bibnamefont {Gough}}, \bibinfo {author} {\bibfnamefont
  {R.}~\bibnamefont {Vafabakhsh}},\ and\ \bibinfo {author} {\bibfnamefont
  {R.}~\bibnamefont {Golestanian}},\ }\bibfield  {title} {\bibinfo {title}
  {Self-motile colloidal particles: From directed propulsion to random walk},\
  }\href@noop {} {\bibfield  {journal} {\bibinfo  {journal} {Phys. Rev. Lett.}\
  }\textbf {\bibinfo {volume} {99}},\ \bibinfo {pages} {048102} (\bibinfo
  {year} {2007})}\BibitemShut {NoStop}%
\bibitem [{\citenamefont {Brown}\ and\ \citenamefont
  {Poon}(2014)}]{brown_poon}%
  \BibitemOpen
  \bibfield  {author} {\bibinfo {author} {\bibfnamefont {A.}~\bibnamefont
  {Brown}}\ and\ \bibinfo {author} {\bibfnamefont {W.}~\bibnamefont {Poon}},\
  }\bibfield  {title} {\bibinfo {title} {Ionic effects in self-propelled
  {P}t-coated {J}anus swimmers},\ }\href@noop {} {\bibfield  {journal}
  {\bibinfo  {journal} {Soft Matter}\ }\textbf {\bibinfo {volume} {10}},\
  \bibinfo {pages} {4016} (\bibinfo {year} {2014})}\BibitemShut {NoStop}%
\bibitem [{\citenamefont {Ebbens}\ \emph {et~al.}(2014)\citenamefont {Ebbens},
  \citenamefont {Gregory}, \citenamefont {Dunderdale}, \citenamefont {Howse},
  \citenamefont {Ibrahim}, \citenamefont {Liverpool},\ and\ \citenamefont
  {Golestanian}}]{ebbens2014electrokinetic}%
  \BibitemOpen
  \bibfield  {author} {\bibinfo {author} {\bibfnamefont {S.}~\bibnamefont
  {Ebbens}}, \bibinfo {author} {\bibfnamefont {D.}~\bibnamefont {Gregory}},
  \bibinfo {author} {\bibfnamefont {G.}~\bibnamefont {Dunderdale}}, \bibinfo
  {author} {\bibfnamefont {J.}~\bibnamefont {Howse}}, \bibinfo {author}
  {\bibfnamefont {Y.}~\bibnamefont {Ibrahim}}, \bibinfo {author} {\bibfnamefont
  {T.}~\bibnamefont {Liverpool}},\ and\ \bibinfo {author} {\bibfnamefont
  {R.}~\bibnamefont {Golestanian}},\ }\bibfield  {title} {\bibinfo {title}
  {Electrokinetic effects in catalytic platinum-insulator {J}anus swimmers},\
  }\href@noop {} {\bibfield  {journal} {\bibinfo  {journal} {EPL}\ }\textbf
  {\bibinfo {volume} {106}},\ \bibinfo {pages} {58003} (\bibinfo {year}
  {2014})}\BibitemShut {NoStop}%
\bibitem [{\citenamefont {Paxton}\ \emph {et~al.}(2004)\citenamefont {Paxton},
  \citenamefont {Kistler}, \citenamefont {Olmeda}, \citenamefont {Sen},
  \citenamefont {St.~Angelo}, \citenamefont {Cao}, \citenamefont {Mallouk},
  \citenamefont {Lammert},\ and\ \citenamefont {Crespi}}]{PaxtonRod}%
  \BibitemOpen
  \bibfield  {author} {\bibinfo {author} {\bibfnamefont {W.~F.}\ \bibnamefont
  {Paxton}}, \bibinfo {author} {\bibfnamefont {K.~C.}\ \bibnamefont {Kistler}},
  \bibinfo {author} {\bibfnamefont {C.~C.}\ \bibnamefont {Olmeda}}, \bibinfo
  {author} {\bibfnamefont {A.}~\bibnamefont {Sen}}, \bibinfo {author}
  {\bibfnamefont {S.~K.}\ \bibnamefont {St.~Angelo}}, \bibinfo {author}
  {\bibfnamefont {Y.}~\bibnamefont {Cao}}, \bibinfo {author} {\bibfnamefont
  {T.~E.}\ \bibnamefont {Mallouk}}, \bibinfo {author} {\bibfnamefont {P.~E.}\
  \bibnamefont {Lammert}},\ and\ \bibinfo {author} {\bibfnamefont {V.~H.}\
  \bibnamefont {Crespi}},\ }\bibfield  {title} {\bibinfo {title} {Catalytic
  nanomotors: Autonomous movement of striped nanorods},\ }\href@noop {}
  {\bibfield  {journal} {\bibinfo  {journal} {J. Am Chem. Soc.}\ }\textbf
  {\bibinfo {volume} {126}},\ \bibinfo {pages} {13424} (\bibinfo {year}
  {2004})}\BibitemShut {NoStop}%
\bibitem [{\citenamefont {de~Buyl}\ and\ \citenamefont
  {Kapral}(2013)}]{de2013phoretic}%
  \BibitemOpen
  \bibfield  {author} {\bibinfo {author} {\bibfnamefont {P.}~\bibnamefont
  {de~Buyl}}\ and\ \bibinfo {author} {\bibfnamefont {R.}~\bibnamefont
  {Kapral}},\ }\bibfield  {title} {\bibinfo {title} {Phoretic self-propulsion:
  a mesoscopic description of reaction dynamics that powers motion},\
  }\href@noop {} {\bibfield  {journal} {\bibinfo  {journal} {Nanoscale}\
  }\textbf {\bibinfo {volume} {5}},\ \bibinfo {pages} {1337} (\bibinfo {year}
  {2013})}\BibitemShut {NoStop}%
\bibitem [{\citenamefont {Buttinoni}\ \emph {et~al.}(2012)\citenamefont
  {Buttinoni}, \citenamefont {Volpe}, \citenamefont {K{\"u}mmel}, \citenamefont
  {Volpe},\ and\ \citenamefont {Bechinger}}]{BechingerJanus}%
  \BibitemOpen
  \bibfield  {author} {\bibinfo {author} {\bibfnamefont {I.}~\bibnamefont
  {Buttinoni}}, \bibinfo {author} {\bibfnamefont {G.}~\bibnamefont {Volpe}},
  \bibinfo {author} {\bibfnamefont {F.}~\bibnamefont {K{\"u}mmel}}, \bibinfo
  {author} {\bibfnamefont {G.}~\bibnamefont {Volpe}},\ and\ \bibinfo {author}
  {\bibfnamefont {C.}~\bibnamefont {Bechinger}},\ }\bibfield  {title} {\bibinfo
  {title} {Active {B}rownian motion tunable by light},\ }\href@noop {}
  {\bibfield  {journal} {\bibinfo  {journal} {J. Phys. Condens. Matter}\
  }\textbf {\bibinfo {volume} {24}},\ \bibinfo {pages} {284129} (\bibinfo
  {year} {2012})}\BibitemShut {NoStop}%
\bibitem [{\citenamefont {Brady}(2011)}]{brady2011}%
  \BibitemOpen
  \bibfield  {author} {\bibinfo {author} {\bibfnamefont {J.~F.}\ \bibnamefont
  {Brady}},\ }\bibfield  {title} {\bibinfo {title} {Particle motion driven by
  solute gradients with application to autonomous motion: continuum and
  colloidal perspectives},\ }\href@noop {} {\bibfield  {journal} {\bibinfo
  {journal} {J. Fluid Mech.}\ }\textbf {\bibinfo {volume} {667}},\ \bibinfo
  {pages} {216} (\bibinfo {year} {2011})}\BibitemShut {NoStop}%
\bibitem [{\citenamefont {J{\"u}licher}\ and\ \citenamefont
  {Prost}(2009)}]{julicher2009}%
  \BibitemOpen
  \bibfield  {author} {\bibinfo {author} {\bibfnamefont {F.}~\bibnamefont
  {J{\"u}licher}}\ and\ \bibinfo {author} {\bibfnamefont {J.}~\bibnamefont
  {Prost}},\ }\bibfield  {title} {\bibinfo {title} {Generic theory of colloidal
  transport},\ }\href@noop {} {\bibfield  {journal} {\bibinfo  {journal} {Eur.
  Phys. J. E}\ }\textbf {\bibinfo {volume} {29}},\ \bibinfo {pages} {27}
  (\bibinfo {year} {2009})}\BibitemShut {NoStop}%
\bibitem [{\citenamefont {Wang}\ and\ \citenamefont
  {Wu}(2014)}]{wang2014bubble}%
  \BibitemOpen
  \bibfield  {author} {\bibinfo {author} {\bibfnamefont {S.}~\bibnamefont
  {Wang}}\ and\ \bibinfo {author} {\bibfnamefont {N.}~\bibnamefont {Wu}},\
  }\bibfield  {title} {\bibinfo {title} {Selecting the swimming mechanisms of
  colloidal particles: bubble propulsion versus self-diffusiophoresis},\
  }\href@noop {} {\bibfield  {journal} {\bibinfo  {journal} {Langmuir}\
  }\textbf {\bibinfo {volume} {30}},\ \bibinfo {pages} {3477} (\bibinfo {year}
  {2014})}\BibitemShut {NoStop}%
\bibitem [{\citenamefont {Zhang}\ \emph {et~al.}(2017)\citenamefont {Zhang},
  \citenamefont {Zheng}, \citenamefont {Cui},\ and\ \citenamefont
  {Silber-Li}}]{zhang2017self}%
  \BibitemOpen
  \bibfield  {author} {\bibinfo {author} {\bibfnamefont {J.}~\bibnamefont
  {Zhang}}, \bibinfo {author} {\bibfnamefont {X.}~\bibnamefont {Zheng}},
  \bibinfo {author} {\bibfnamefont {H.}~\bibnamefont {Cui}},\ and\ \bibinfo
  {author} {\bibfnamefont {Z.}~\bibnamefont {Silber-Li}},\ }\bibfield  {title}
  {\bibinfo {title} {The self-propulsion of the spherical {P}t--{S}i{O}$_2$
  {J}anus micro-motor},\ }\href@noop {} {\bibfield  {journal} {\bibinfo
  {journal} {Micromachines}\ }\textbf {\bibinfo {volume} {8}},\ \bibinfo
  {pages} {123} (\bibinfo {year} {2017})}\BibitemShut {NoStop}%
\bibitem [{\citenamefont {Herminghaus}\ \emph {et~al.}(2014)\citenamefont
  {Herminghaus}, \citenamefont {Maass}, \citenamefont {KrÃŒger},
  \citenamefont {Thutupalli}, \citenamefont {Goehring},\ and\ \citenamefont
  {Bahr}}]{Herminghaus14}%
  \BibitemOpen
  \bibfield  {author} {\bibinfo {author} {\bibfnamefont {S.}~\bibnamefont
  {Herminghaus}}, \bibinfo {author} {\bibfnamefont {C.~C.}\ \bibnamefont
  {Maass}}, \bibinfo {author} {\bibfnamefont {C.}~\bibnamefont {KrÃŒger}},
  \bibinfo {author} {\bibfnamefont {S.}~\bibnamefont {Thutupalli}}, \bibinfo
  {author} {\bibfnamefont {L.}~\bibnamefont {Goehring}},\ and\ \bibinfo
  {author} {\bibfnamefont {C.}~\bibnamefont {Bahr}},\ }\bibfield  {title}
  {\bibinfo {title} {Interfacial mechanisms in active emulsions},\ }\href@noop
  {} {\bibfield  {journal} {\bibinfo  {journal} {Soft Matter}\ }\textbf
  {\bibinfo {volume} {10}},\ \bibinfo {pages} {7008} (\bibinfo {year}
  {2014})}\BibitemShut {NoStop}%
\bibitem [{pro(2008)}]{propellant}%
  \BibitemOpen
  \bibfield  {title} {\bibinfo {title} {``{R}ocket {P}ropellants''},\ }in\
  \href@noop {} {\emph {\bibinfo {booktitle} {Van Nostrand's Encyclopedia of
  Science, Volume 1}}},\ \bibinfo {editor} {edited by\ \bibinfo {editor}
  {\bibfnamefont {G.~D.}\ \bibnamefont {Considine}}\ and\ \bibinfo {editor}
  {\bibfnamefont {P.~H.}\ \bibnamefont {Kulik}}}\ (\bibinfo  {publisher}
  {Wiley-Interscience},\ \bibinfo {year} {2008})\ \bibinfo {edition} {10th}\
  ed.,\ pp.\ \bibinfo {pages} {4545--4548}\BibitemShut {NoStop}%
\bibitem [{\citenamefont {Cichocki}\ and\ \citenamefont
  {Felderhof}(2000)}]{Felderhof2000long}%
  \BibitemOpen
  \bibfield  {author} {\bibinfo {author} {\bibfnamefont {B.}~\bibnamefont
  {Cichocki}}\ and\ \bibinfo {author} {\bibfnamefont {B.}~\bibnamefont
  {Felderhof}},\ }\bibfield  {title} {\bibinfo {title} {Long-time tails in the
  solid-body motion of a sphere immersed in a suspension},\ }\href@noop {}
  {\bibfield  {journal} {\bibinfo  {journal} {Physical Review E}\ }\textbf
  {\bibinfo {volume} {62}},\ \bibinfo {pages} {5383} (\bibinfo {year}
  {2000})}\BibitemShut {NoStop}%
\bibitem [{\citenamefont {Hoogerbrugge}\ and\ \citenamefont
  {Koelman}(1992)}]{hoogerbrugge1992dpd}%
  \BibitemOpen
  \bibfield  {author} {\bibinfo {author} {\bibfnamefont {P.}~\bibnamefont
  {Hoogerbrugge}}\ and\ \bibinfo {author} {\bibfnamefont {J.}~\bibnamefont
  {Koelman}},\ }\bibfield  {title} {\bibinfo {title} {Simulating microscopic
  hydrodynamic phenomena with dissipative particle dynamics},\ }\href@noop {}
  {\bibfield  {journal} {\bibinfo  {journal} {EPL}\ }\textbf {\bibinfo {volume}
  {19}},\ \bibinfo {pages} {155} (\bibinfo {year} {1992})}\BibitemShut
  {NoStop}%
\bibitem [{\citenamefont {Groot}\ and\ \citenamefont
  {Warren}(1997)}]{groot1997dissipative}%
  \BibitemOpen
  \bibfield  {author} {\bibinfo {author} {\bibfnamefont {R.~D.}\ \bibnamefont
  {Groot}}\ and\ \bibinfo {author} {\bibfnamefont {P.~B.}\ \bibnamefont
  {Warren}},\ }\bibfield  {title} {\bibinfo {title} {Dissipative particle
  dynamics: Bridging the gap between atomistic and mesoscopic simulation},\
  }\href@noop {} {\bibfield  {journal} {\bibinfo  {journal} {J. Chem. Phys.}\
  }\textbf {\bibinfo {volume} {107}},\ \bibinfo {pages} {4423} (\bibinfo {year}
  {1997})}\BibitemShut {NoStop}%
\bibitem [{\citenamefont {Espanol}\ and\ \citenamefont
  {Warren}(1995)}]{espanol1995statistical}%
  \BibitemOpen
  \bibfield  {author} {\bibinfo {author} {\bibfnamefont {P.}~\bibnamefont
  {Espanol}}\ and\ \bibinfo {author} {\bibfnamefont {P.~B.}\ \bibnamefont
  {Warren}},\ }\bibfield  {title} {\bibinfo {title} {Statistical mechanics of
  dissipative particle dynamics},\ }\href@noop {} {\bibfield  {journal}
  {\bibinfo  {journal} {EPL}\ }\textbf {\bibinfo {volume} {30}},\ \bibinfo
  {pages} {191} (\bibinfo {year} {1995})}\BibitemShut {NoStop}%
\bibitem [{\citenamefont {Easton}\ \emph {et~al.}(1952)\citenamefont {Easton},
  \citenamefont {Mitchell},\ and\ \citenamefont {Wynne-Jones}}]{h2o2density}%
  \BibitemOpen
  \bibfield  {author} {\bibinfo {author} {\bibfnamefont {M.}~\bibnamefont
  {Easton}}, \bibinfo {author} {\bibfnamefont {A.}~\bibnamefont {Mitchell}},\
  and\ \bibinfo {author} {\bibfnamefont {W.}~\bibnamefont {Wynne-Jones}},\
  }\bibfield  {title} {\bibinfo {title} {The behaviour of mixtures of hydrogen
  peroxide and water},\ }\href@noop {} {\bibfield  {journal} {\bibinfo
  {journal} {Trans. Faraday Soc.}\ }\textbf {\bibinfo {volume} {48}},\ \bibinfo
  {pages} {796} (\bibinfo {year} {1952})}\BibitemShut {NoStop}%
\bibitem [{\citenamefont {Hasimoto}(1959)}]{hasimoto1959periodic}%
  \BibitemOpen
  \bibfield  {author} {\bibinfo {author} {\bibfnamefont {H.}~\bibnamefont
  {Hasimoto}},\ }\bibfield  {title} {\bibinfo {title} {On the periodic
  fundamental solutions of the stokes equations and their application to
  viscous flow past a cubic array of spheres},\ }\href@noop {} {\bibfield
  {journal} {\bibinfo  {journal} {J. Fluid Mech.}\ }\textbf {\bibinfo {volume}
  {5}},\ \bibinfo {pages} {317} (\bibinfo {year} {1959})}\BibitemShut {NoStop}%
\end{thebibliography}%

\clearpage
\section*{\label{sec:level2} Supporting Information}

\begin{figure}[h!]
  \centering
  \begin{tabular}{@{}c@{}}
     \small (a) 
     \includegraphics[width=.8\linewidth]{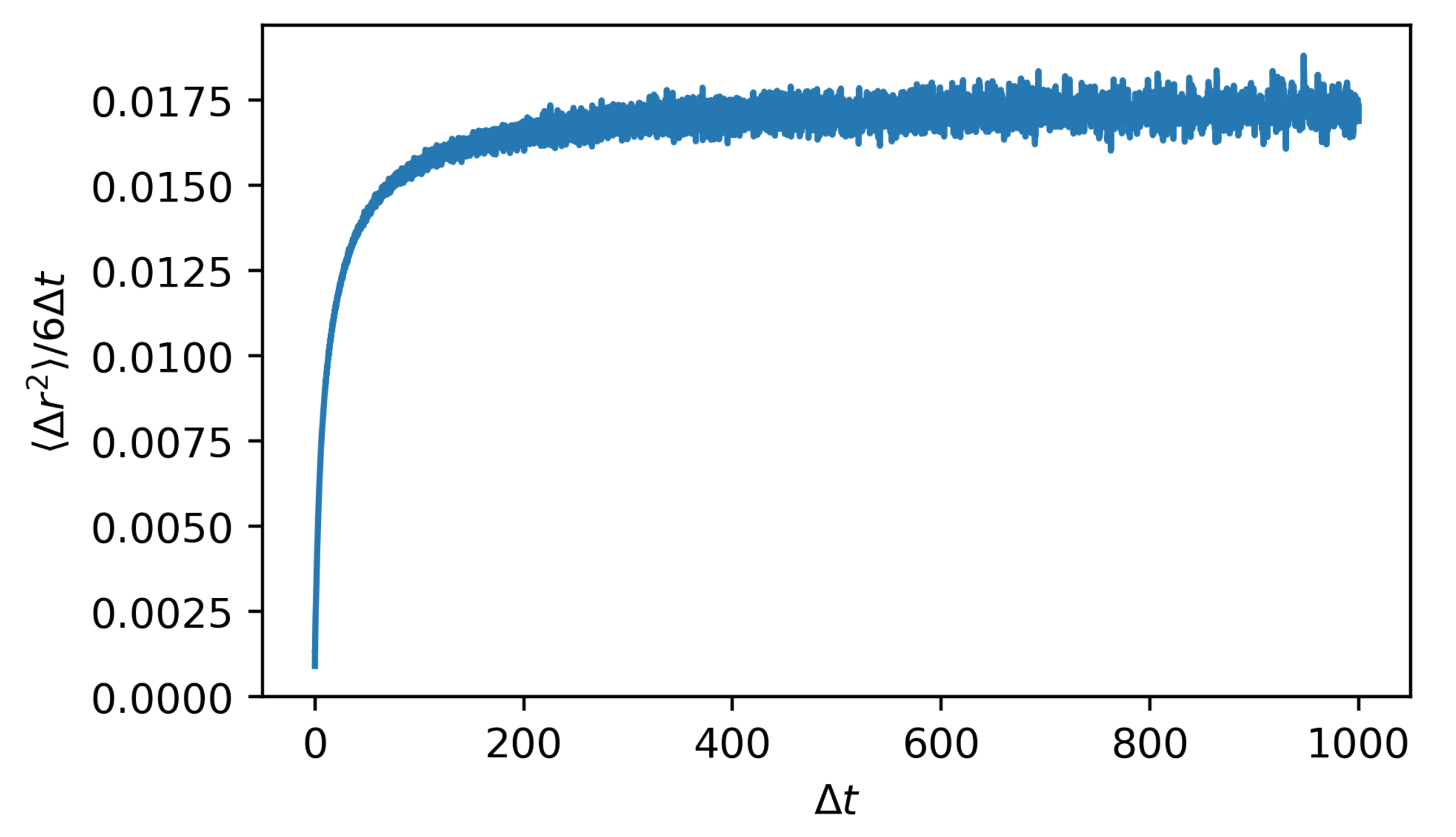} \\[\abovecaptionskip]
  \end{tabular}
  \begin{tabular}{@{}c@{}}
   \small (b)
   \includegraphics[width=.75\linewidth]{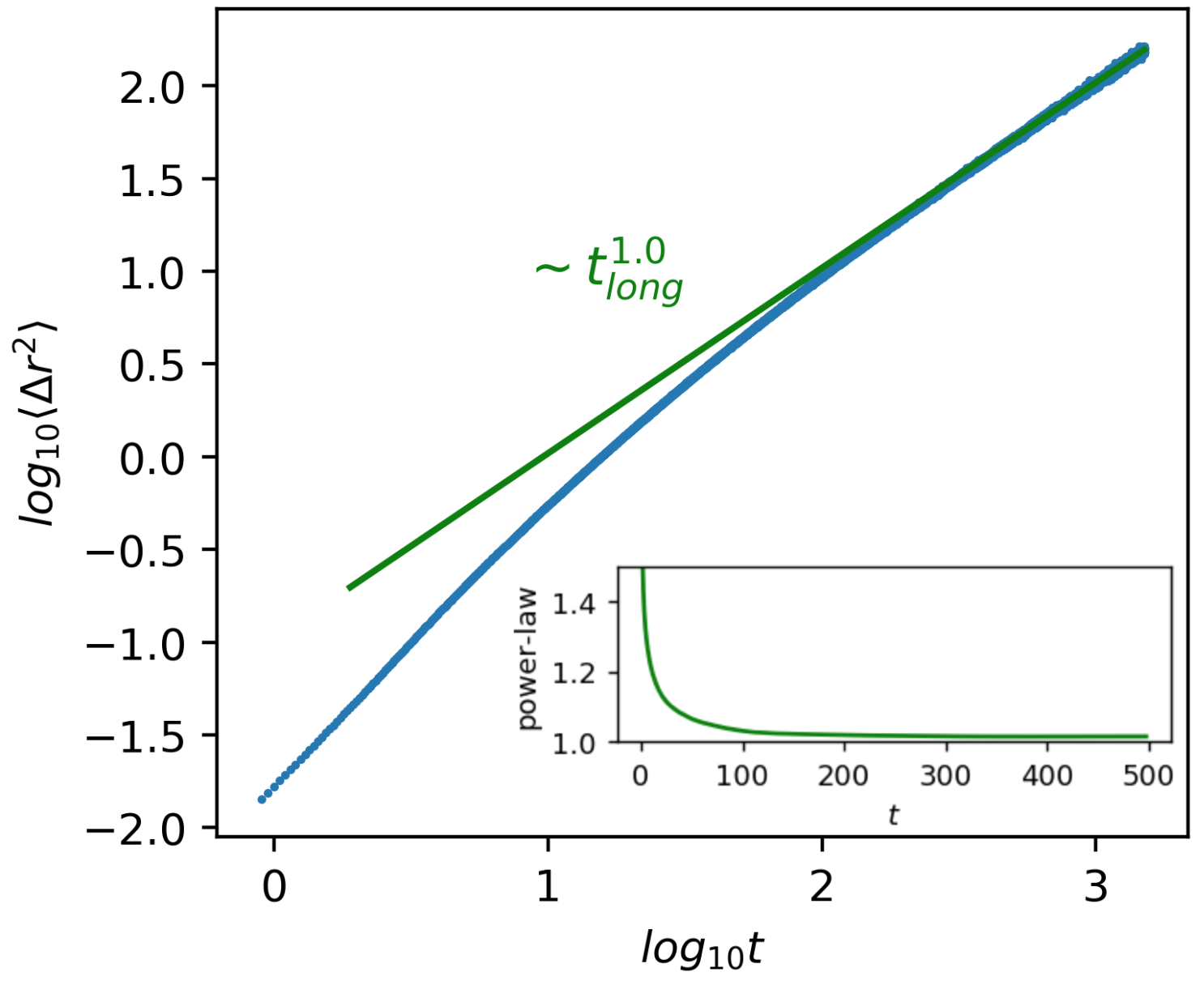} \\[\abovecaptionskip]
  \end{tabular} 
  \caption{\label{fig:msdondt1000} (a) Example of the mean square displacement of a non-active colloid for the case where the ratio of the box length $L$ to the colloidal radius $R $ equals $L/R =7.5$. 
(b) A Log-log plot of the colloidal mean-squared displacement versus time. As is clear from the figure, normal diffusive behavior (exponent $= 1$) is observed after some 200 time units. The inset shows the time dependence of the effective exponent, also showing the approach to the diffusive regime at long times. 
}
\end{figure}
 \begin{figure}[t!]
  \centering
  \begin{tabular}{@{}c@{}}
     \small (a) 
     \includegraphics[width=.9\linewidth]{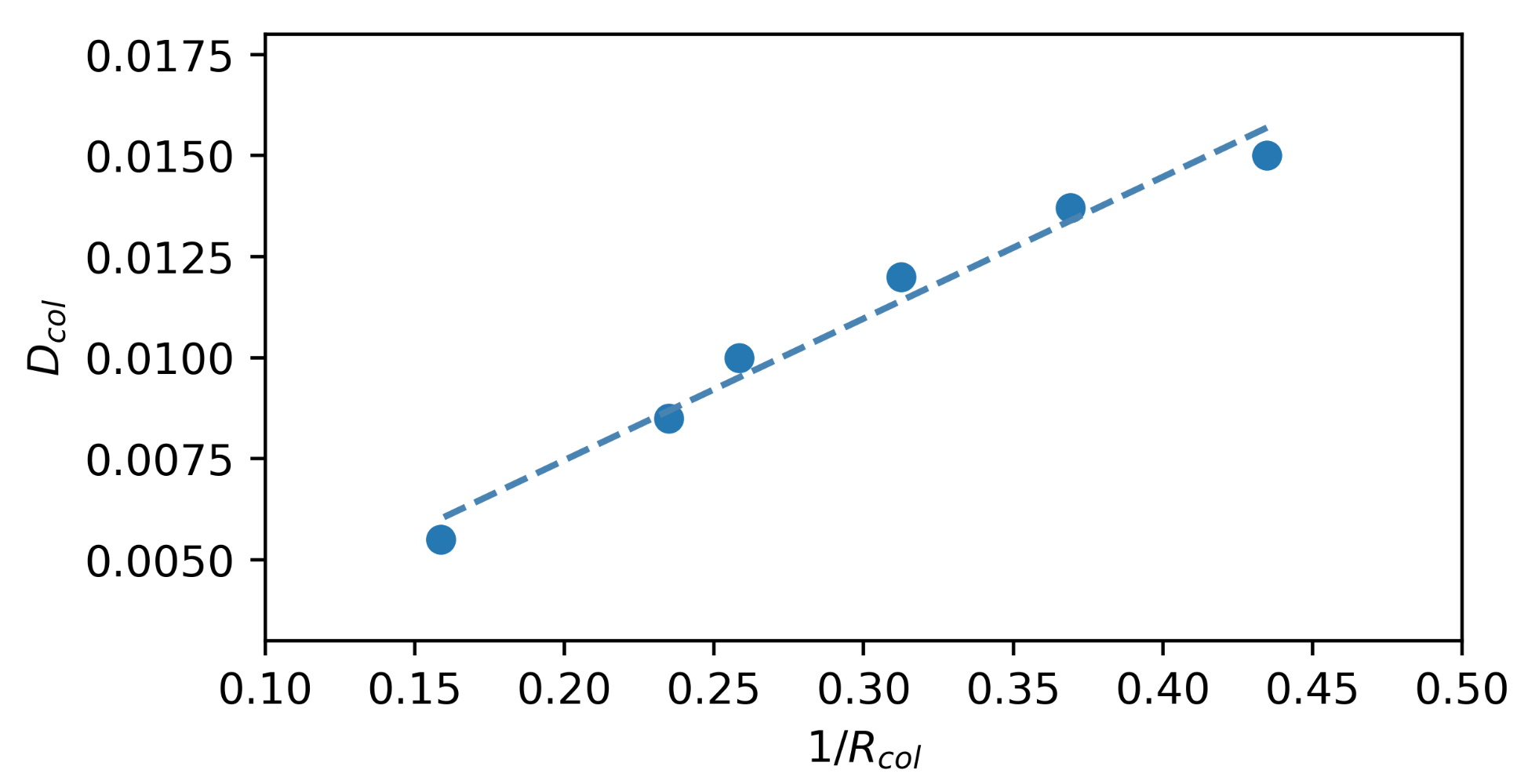} \\[\abovecaptionskip]
  \end{tabular}
  \begin{tabular}{@{}c@{}}
   \small (b)
   \includegraphics[width=.9\linewidth]{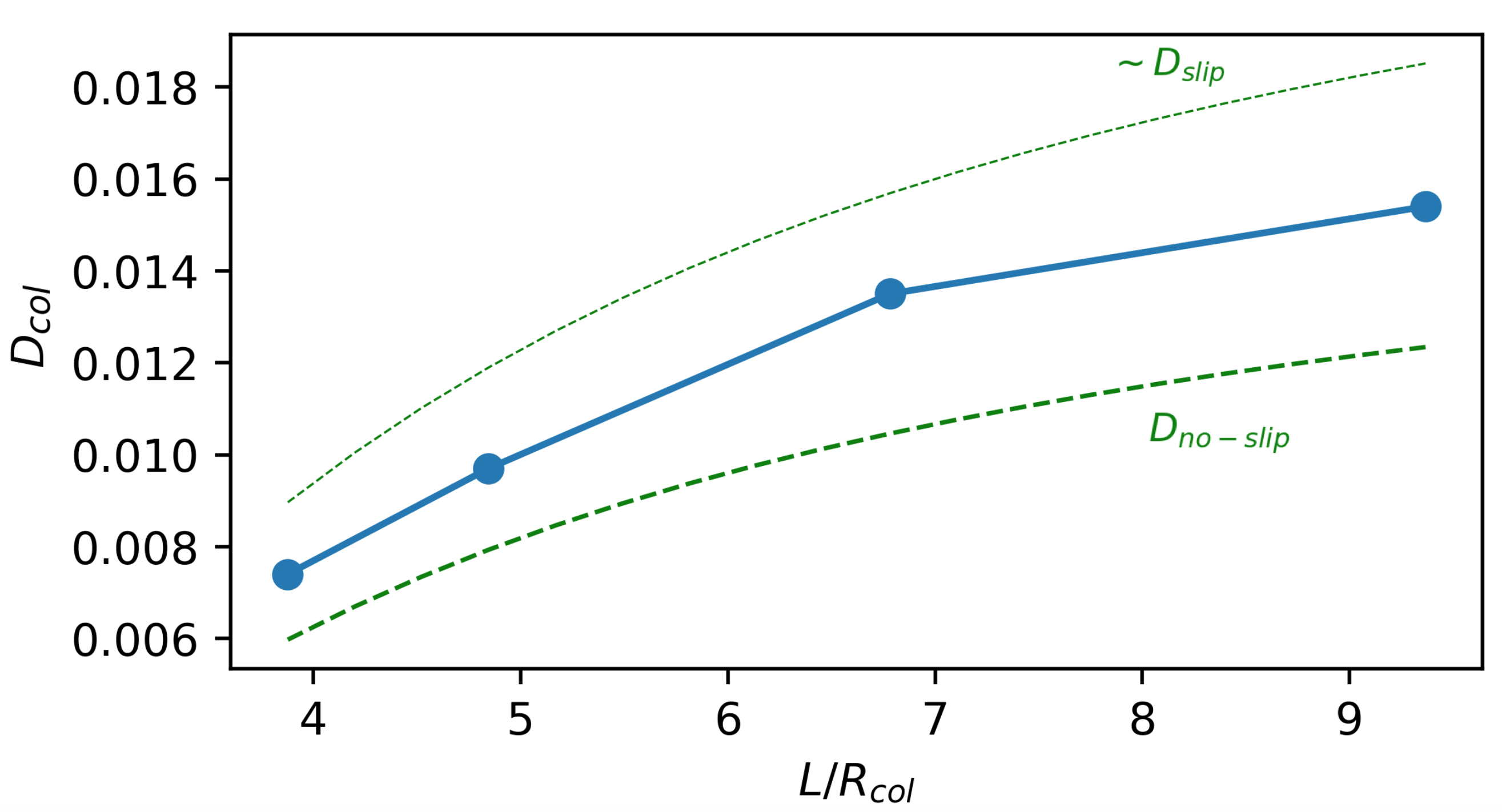} \\[\abovecaptionskip]
    
  \end{tabular}
  \caption{(a) The diffusion coefficient of the colloid as function of the inverse of the colloid radius, $1/R $, whilst keeping the ratio $L/R =7.5$. The curve indicates that $D \sim 1/R $, as expected from Stokes-Einstein relation. (b) The diffusion coefficient calculated for various ratios of box sizes to colloid radius (solid circles, solid line is a guide to the eye). The dash lines depict the non-slip (lower) and estimated slip (higher) limits. The colloid surface behaviour is approximately between slippery and non-slippery boundary conditions. }\label{fig:stoke}
\end{figure}

 
{\bf Passive colloid:} To validate our model, we first simulated the Brownian diffusion of a passive colloid in the DPD fluid with periodic boundary conditions. This allows us to estimate the effective hydrodynamic radius of the colloid. 
Due to hydrodynamic interactions, the diffusion coefficient of a colloidal particle depends strongly on the size of the periodic box, $L$. To account for this finite-size effect, we carried out simulations with various periodic box sizes and also, to check consistency, simulations of colloids with different radii in a fixed periodic box. In these simulations we measured the mean squared displacement (MSD) of the colloid in the fluid. For sufficiently long times $t$, the MSD, $\langle \Delta r^2 \rangle= \langle [r(t + \Delta t) -r(t)]^2\rangle$ approaches  $6D \Delta t$, where $D $ is the colloid diffusion coefficient. This is demonstrated in Fig.~\ref{fig:msdondt1000}a and b, showing the MSD as a function of the time from simulations of a particle of radius $R =2.3$ (corresponding to 1.5nm in physical units) and $L=7.5R $. We estimate $D $ from
\begin{eqnarray*}
D  =  \left.\frac{\langle \Delta r^2 \rangle}{6\Delta t}\right\vert_{\Delta t\rightarrow\infty}.
\end{eqnarray*} 
The MSD analysis exhibits a typical transition from ballistic to diffusive behaviour. 
A comparison with the Stokes-Einstein relation is also provided in Fig.~\ref{fig:stoke}a, showing that $D $ scales inversely with $R $. We stress that as we do not impose non-slip boundary conditions in the simulations, the diffusion coefficient of the colloid is expected to be higher than the non-slip limit ($D^\infty=\frac{kT}{6\pi \eta R })$. To allow comparison, we also need to account for finite-system-size effect, which can be done using  the Hasimoto correction~\cite{hasimoto1959periodic}:
\begin{eqnarray}
D_{\rm no-slip} \sim D^\infty \left( 1-\frac{2.8373R }{L} +\frac{4\pi R ^3}{3L^3}\right).
\end{eqnarray} 
Fig.~\ref{fig:stoke}b demonstrates that the diffusion coefficient of a colloid with radius $R $=3.2 (corresponding to $\sim \SI{2}{\nano\meter}$) is located approximately halfway between the slip and no-slip limits ($D_{\rm slip} =  \frac{3}{2}D_{\rm no-slip}$).
Using the Groot and Warren expression for the kinematic viscosity~\cite{groot1997dissipative}: $\nu\approx 45k_bT/(4\pi \gamma \rho r_c^3) +2\pi\gamma \rho r_c^5/1575$, we find the dynamic viscosity 
$\eta=0.96$ for the DPD fluid parameters used. 
We note that the Stokes radius of the colloid can vary slightly with the surface density of `frozen' DPD particles.

\end{document}